\documentclass[usenatbib]{mn2e}
\usepackage[dvips]{graphicx}
\usepackage{multirow}
\usepackage{amsmath}

\def\CN2{\mbox{$C_N^2$}}
\def\tauO{\mbox{$\tau_{0}$}}
\def\thetaO{\mbox{$\theta_{0}$}}

\title[Forecast of surface layer parameters at C. Paranal]{Forecast of surface layer meteorological parameters at Cerro Paranal with a mesoscale atmospherical model.}
\author[F. Lascaux et al.]{F. Lascaux$^{1}$\thanks{E-mail:
     lascaux@arcetri.astro.it; masciadri@arcetri.astro.it}, E. Masciadri$^1$\footnotemark[1] and L. Fini$^1$ \\
$^1$INAF Osservatorio Astrofisico di Arcetri, Largo Enrico Fermi 5, I-501 25 Florence, Italy}

\begin{document}
\newcommand{\cn}{$C_N^2$}
\label{firstpage}
\date{Accepted 2014 ??? ??, Received 2014 ??? ??; in original form
2014 ??? ??}
\pagerange{\pageref{firstpage}--\pageref{lastpage}}
\pubyear{2013}
\maketitle
\begin{abstract}
This article aims at proving the feasibility of the forecast of all the most 
relevant classical atmospherical parameters for astronomical applications (wind speed and direction, temperature) 
above the ESO ground-base site of Cerro Paranal with a mesoscale atmospherical model called Meso-Nh. 
In a precedent paper we have preliminarily treated the model performances obtained in reconstructing some key atmospherical parameters
in the surface layer 0-30~m studying the bias and the RMSE on a statistical sample of 20 nights. Results were very encouraging 
and it appeared therefore mandatory to confirm such a good result on a much richer statistical sample.
In this paper, the study was extended to a total sample of 129 nights between 2007 and 2011 distributed in different parts of the solar year.
This large sample made our analysis more robust and definitive in terms of the model performances and permitted us to confirm
the excellent performances of the model. Besides, we present an independent analysis of the model performances 
using the method of the contingency tables. Such a method permitted us to provide complementary key informations with respect to the bias and the RMSE
particularly useful for an operational implementation of a forecast system.
\end{abstract}
\section{Introduction}
This paper is  part of a general study about the feasibility of the forecast of meteorological parameters and optical turbulence 
at ESO sites (Cerro Paranal and Cerro Armazones) in the framework of the MOSE project (MOdeling ESO Sites). 
The MOSE project is presented extensively in a previous paper \citep{masciadri2013}.
We only recall here that the MOSE project aims at proving the feasibility of the forecast of all the most relevant classical atmospherical parameters
for astronomical applications (wind speed intensity and direction, temperature, relative humidity) and the optical turbulence OT ($\CN2$ profiles)
with the integrated astro-climatic parameters derived from the $\CN2$ i.e. the seeing ($\varepsilon$),
the isoplanatic angle ($\thetaO$), the wavefront coherence time ($\tauO$) above the two ESO sites of Cerro Paranal
(site of the Very Large Telescope - VLT) and Cerro Armazones (site selected for the European Extremely Large Telescope - E-ELT).\\
The final outcome of the project is to investigate the opportunity to implement an automatic system for the
forecast of these parameters at the VLT Observatory at Cerro Paranal and at the E-ELT Observatory at Cerro Armazones.\\ 
In a previous paper \citep{lascaux2013}, we presented results of a deep analysis of the bias and RMSE between observations and 
model outputs of absolute temperature, wind speed and direction above both astronomical sites (Paranal and Armazones). 
These statistical operators provide fundamental informations on systematic and statistical model errors. 
This statistical study was performed on a sample of 20 nights, the same for each site. 
A part of the nights is coming from the PAR2007 campaign \citep{dali10}, and the sample was completed with nights with available observations from the same period
 of the year. 
In the same study we also provided a detailed analysis on the model performances on individual nights. 
The most important results we obtained can be summarized as in the following. 
The model showed a very good score of success in terms of bias and RMSE for absolute 
temperature (median bias in the [0.03,0.64]$^{\circ}$C range and median RMSE in the [0.64,0.93]$^{\circ}$C range) 
and wind direction (median bias in the [-1.01,-8.55]$^{\circ}$ range and median RMSE in the [30,41]$^{\circ}$ range for a wind speed threshold of 
2~$m\ s^{-1}$ equivalent to a RMSE$_{relative}$ in the [17,23]$\%$ range) when it is used in grid-nesting configuration and the horizontal resolution 
of the innermost model domain is 500~m. 
To obtain equivalent satisfactory results for the wind speed 
we need to use a horizontal resolution of the innermost model domain equal to 100~m. Under this condition we obtained a median bias within 
0.93~$m\ s^{-1}$ and a median RMSE within 2.18~$m\ s^{-1}$. These results were, in conclusion, very promising.\\ 

In this paper the preliminary analysis done by \citet{lascaux2013} was extended to a sample of 129 nights uniformly distributed between 2007 and 2011. 
The analysis is performed for the Cerro Paranal case. 
It was possible to have access to homogeneous measurements distributed over such a long period of time, only at this site\footnote{At Cerro Armazones during this period 
part of measurements have been done with TMT instrumentation and part with ESO one with sensors located at a different number of levels and heights above the ground. 
This lack of homogeneity in measurements did not permit us to perform the analysis at Armazones using the same procedure. 
We therefore have decided in this paper to focus our attention only on Cerro Paranal.}.
All the 129 nights belong to different periods of the year, and not only to summer (like for the sample of 20 nights from \cite{lascaux2013}).
This permits us to analyze definitely the robustness of the model on a rich statistical sample, independently from the period of the year considered. 
All the measurements, at Cerro Paranal, are part of the VLT Astronomical Site Monitor \citep{sandrock99}.\\ 

Besides that, on the same sample of nights, we present a complementary statistical analysis based on a different approach. 
Indeed, it is worth to highlight that bias and RMSE, in spite to be fundamental statistical operators providing key informations on model performances, do not answer to the 
necessary information one would like to have in terms of model performances \citep{thornes2001,nurmi2003,jolliffe2003}. 
A key rule in the complex art of the model prediction estimation is that it makes 
no sense to quantify the model performance with one parameter only or one method only. 
To investigate the quality of a model prediction, a method widely used in physics of the atmosphere 
as well as in other fields such as economy and medicine, consists in constructing and analyzing contingency tables. 
From these tables one can derive a number of different parameters that describe the quality of the model performances. \\
A contingency table allows for the analysis of the relationship between two or more categorical variables. 
One of the first classic examples of a forecast verification using contingency tables is the tornado forecast from \citet{finley1884}.
A contingency table is a table with $n{\times}n$ entries that displays the distribution of modeled outputs and observations in terms of frequencies 
or relative frequencies.
 Here the variables we considered are 
the observations of different categories of temperature, wind speed, wind direction, near the surface, and the respective reconstructed parameters 
by the Meso-NH model.  \\ 

In this part of Chile the weather is particularly dry, the relative humidity (RH) in the surface layer is typically below 30$\%$.
The frequency of the nights with a relative humidity above 80$\%$, 
level at which the dome of the telescope has to be closed and observations are not allowed, is definitely very low
(typically less than 5 nights for year). In all the other nights the RH does not really represent a critical issue in terms of scheduling of observations 
because the RH value is not high enough to affect the observations. The study of the relative humidity close to the ground is therefore not very critical and important 
for telescopes in this region. Also it has been observed in occasion of the TMT site characterization in this region (T. Travouillon, private communication),
that the reliability of the relative humidity measurements, when values are so low,
can hardly be assured. This makes difficult (and mostly useless) to investigate the model performances
in reconstructing the RH on the same sample of nights selected for the analysis of the other parameters. A more interesting analysis
might be to check the model performances in reconstructing the RH values in those few nights in each year where RH is higher than the threshold
imposing to close the dome of the telescope.  In Annex we report satisfactory results obtained in this sense.\\ 

In Section \ref{sec2} we describe the model configuration defined for this study. 
In Section \ref{sec3} we report the bias and RMSE obtained for the new sample of 129 nights, at Cerro Paranal.
In Section \ref{sec4} we define what is a contingency table for different discretized values $\it{n}$ of observations and predictions 
and which parameters qualifying the model behavior can be retrieved from them. 
In Section \ref{sec:climato} we describe the criteria we used to identify the thresholds separating the $\it{n}$ categories. 
In Sections \ref{sec6}, \ref{sec7} and \ref{sec8} we provide the results in terms of model performances
(reconstruction of the absolute temperature, the wind speed and direction in the 0-30~m range).
Finally conclusions are drawn in Section \ref{sec9}.
\section{Brief overview of the numerical configuration}       
\label{sec2}          
\subsection{Sample selection}
The 129 nights have been selected starting with the 20 nights of \citet{lascaux2013}. 
Those 20 nights were all concentrated in November and December 2007 (the reader can refer to \citet{lascaux2013} to understand how this first sample was constructed).
When results obtained with the sample of 20 nights indicated a very promising model behavior, to have a better evaluation of the model performances in statistical terms, it was decided to extend this sample to a more consistent one.
Two different periods, distant in time, were arbitrarily chosen (the only constraint was the availability of continuous measurements): 1/from January 2007 to December 2007 and 2/from June 2010 to May 2011.
We tried to select nights that were regularly spaced in time, as much as possible, in order to cover all the four seasons (which are characterized by 
slightly different meteorological behaviours).
The result was 55 nights in 2007 (the 20 ones mentioned above, plus around 3 nights per month), and 74 nights in 2010-2011 (around 6 nights per month).
No specific criterium has been used on the goodness of the weather conditions. In other words, nights include good and/or bad weather conditions therefore no specific bias affects the sample. All these nights were characterized by different conditions of wind speed, temperature and relative humidity (see discussion in Section \ref{sec3}).
\subsection{Model configuration}
All the numerical simulations of the nights (129 in total) presented in this study have been performed with the mesoscale numerical weather model
 Meso-NH\footnote{$http$:$//mesonh.aero.obs$-$mip.fr/mesonh/$} \citep{lafore98} and the Astro-Meso-NH code for the optical turbulence \citep{masciadri1999}.
The model has been developed by the Centre National des Recherches M\'et\'eorologiques (CNRM) and Laboratoire d'A\'erologie (LA)
from Universit\'e Paul Sabatier (Toulouse).
The Meso-NH model can simulate the temporal evolution of three-dimensional meteorological
parameters over a selected finite area of the globe.
We refer the reader to \citet{masciadri2013}, Sec.~3.3, for the general model configuration and the physical packages used for this study.
We just recall here that we used the grid-nesting technique \citep{stein00}, that consists in using different imbricated domains
of the Digital Elevation Models (DEM i.e orography) extended on smaller and smaller surfaces, with increasing horizontal
resolution but with the same vertical grid.
In this study we use two different configurations.
The first grid-nesting configuration employed three domains (Table~\ref{tab:config}) and the innermost resolution is ${\Delta}X$~=~500~m.
The second configuration is made of four imbricated domains, the first same three as the previous configuration, plus one
centered at Cerro Paranal site, with a horizontal resolution of ${\Delta}X$~=~100~m (all domains of 
Table~\ref{tab:config}).
Along the z-axis we have 62 levels distributed as follows: a first vertical grid point equal to 5~m,
a logarithmic stretching of 20~$\%$ up to 3.5~km above the ground, and an almost constant vertical grid size of $\sim$600~m up to 23.8~km.\\
All the simulations done for the analyses discussed in this paper, were initialized the day before at 18~UT
and forced every 6 hours with the analyses from the ECMWF (European Centre for Medium-Range Weather Forecasts),
and finished at 09~UT (05 LT) of the simulated day (for a total duration of 15 hours).
The reader can refer to \citet{lascaux2013} to have more information about the site coordinates and ground altitude.\\
\begin{table*}
\centering
\caption{Meso-NH model configurations. In the second column the  horizontal resolution ${\Delta}X$, in the third column the number of grid points and in the fourth column the horizontal surface covered by the model domain.}
\begin{tabular}{cccc}
\hline
\multirow{2}{*}{Domain} & \multirow{2}{*}{${\Delta}X$ (km)} & \multirow{2}{*}{Grid Points}    & Domain size \\
                        &                                 &                                 & (km)        \\ 
\hline
Domain 1     & 10             &  80$\times$80  &  800$\times$800       \\
Domain 2     &  2.5           &  64$\times$64  &  160$\times$160       \\
Domain 3     &  0.5           & 150$\times$100 &   75$\times$50        \\
Domain 4     &  0.1           & 100$\times$100 &   10$\times$10        \\
\hline
\end{tabular}
\label{tab:config}
\end{table*}
Before computing bias and RMSE, and before constructing the contingency tables, all the investigated meteorological parameters (measurements and Meso-NH outputs) have been 
first averaged with a moving average of 1 hour to cut the high frequencies and then resampled over a 20-minute interval. The moving average of one hour has been selected because astronomers are in reality more interested in the trend of the prediction. It has been observed that the moving average is more efficient to identify the model and measurements trends than a simple re-sampling. The high frequency variability on a time scale of 5 or 10 minutes is less relevant and useless to predict because the scheduling of scientific programs could not be tuned with such high frequency. Astronomers are interested in identifying if the trend of a parameter is increasing, decreasing or is stationary to be able to take a decision in changing a modality of observation or a scientific program. The selection of the interval of 20 minutes for the re-sampling of measurements is justified 
by the fact that this is more or less the effective time necessary to switch from a modality of observation to another one. The values of one hour and 20 minutes have been selected in agreement with ESO staff.
In the analysis presented in this paper we will consider the model abilities in reconstructing the 
atmospherical parameters during nighttime only\footnote{To perform daytime predictions, a different model configuration
should be selected.}.
\section{Statistical analysis}
\label{sec3}
\begin{figure*}
\begin{center}
\begin{tabular}{c}
\includegraphics[width=0.53\textwidth]{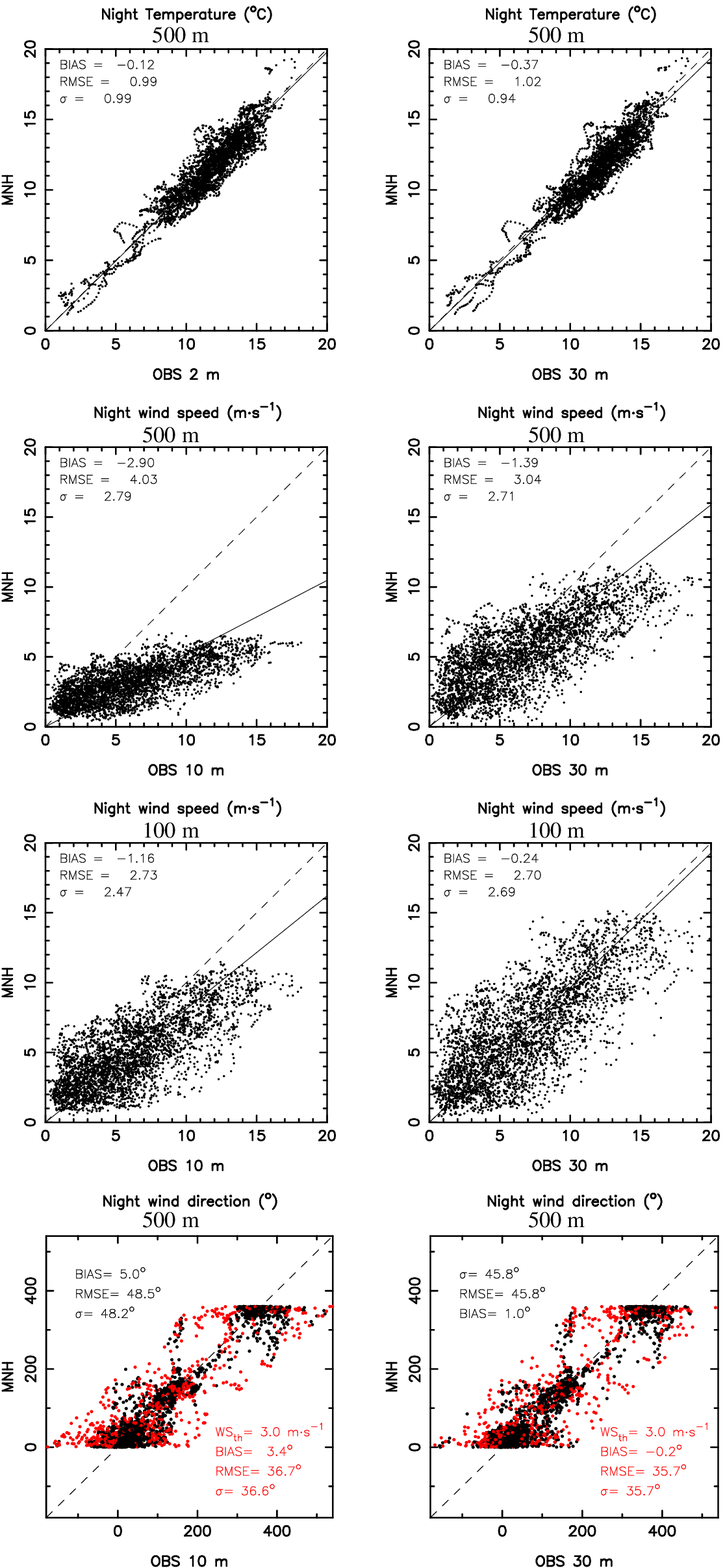}\\
\end{tabular}
\end{center}
\caption{\label{fig:bias_rmse_all} Scattered plots of the absolute temperature (top row), of the wind speed with the $\Delta$X~=~500~m configuration 
(second row), of the wind speed with the $\Delta$X~=~100~m configuration (third row) and of the wind direction (bottom row).
A moving average (1 hour) have been applied on both measurements and Meso-NH outputs (cf. text). Each point represents an average of 20 minutes. The thin black line 
is the regression line passing by the origin.
The red dots correspond to points for which the observed wind was inferior to $WS_{th}$~=~3~m$\cdot$s$^{-1}$.
The corresponding values of bias and RMSE of the wind direction 
for the sample without these dots, are reported in red in the bottom right of the wind direction scattered plots.
The values of the biases and the RMSE are reported in Table~\ref{tab:br_129n_par}.}
\end{figure*}

In this section we estimate the statistical model reliability using three statistical operators: the bias, the root mean square error (RMSE) and the bias-corrected RMSE.
\begin{equation}
BIAS = \sum_{i=1}^{N}\frac{({\Delta}_i)}{N}
\label{eq:bias}
\end{equation}
\begin{equation}
RMSE = \sqrt{\sum_{i=1}^{N}\frac{({\Delta}_i)^2}{N}}
\label{eq:rmse}
\end{equation}
with ${\Delta}_i=Y_i-X_i$
where $X_i$ are the individual observations, $Y_i$ the individual simulations parameters calculated at the same time and $N$ is
the number of times for which a couple ($X_i$,$Y_i$) is available with both $X_i$ and $Y_i$ different from zero.
Because the wind direction is a circular variable, we define ${\Delta}_i$ for the wind direction as:
\begin{equation}
\Delta_i = \begin{cases}
Y_i-X_i     & \text{if $\left | Y_i-X_i  \right | \leq 180{^{\circ}}$} \\
Y_i-X_i-360{^{\circ}} & \text{if $Y_i-X_i >  180{^{\circ}}$} \\
Y_i-X_i+360{^{\circ}} & \text{if $Y_i-X_i < -180{^{\circ}}$}
\end{cases}
\label{eq:deltai}
\end{equation}
From the bias and the RMSE we deduce the bias-corrected RMSE ($\sigma$):
\begin{equation}
\sigma = \sqrt{RMSE^{2}-BIAS^{2}}\\
\label{eq:bcr}
\end{equation}
which represents the error of the model once the systematic error (the bias) is removed.
For the wind direction we add another operator, the relative RMSE:
\begin{equation}
RMSE_{rel} = \frac{RMSE}{180} \times 100 \% \\
\label{eq:relrmse}
\end{equation}
with RMSE expressed in degrees.
This is justified because the wind direction is a circular variable and the maximum error is obtained when RMSE is equal to 180$^{\circ}$.\\
Figure~\ref{fig:bias_rmse_all} shows the scattered plots for the temperature, the wind speed (with both $\Delta$X~=~500~m and $\Delta$X~=~100~m configurations) and the 
wind direction (with and without filtering the lowest wind speed).
The first thing to notice is the total extent of the observed values for all the meteorological parameters. 
The temperature goes from around 0$^{\circ}$C to 20$^{\circ}$C, the wind speed reaches 20 m$\cdot$s$^{-1}$ and the wind direction covers the whole [0-360]$^{\circ}$ range.
This is in agreement with the minimum and the maximum abserved values during a whole year at the site, as is reported on the Astoclimatology Website of 
Paranal\footnote{http://www.eso.org/gen-fac/pubs/astclim/paranal}.
This assures us that all the possible different meteorological conditions were encountered for the 129 simulated nights.
\begin{table*}
 \centering
 \caption{Near surface bias, RMSE, bias-corrected RMSE $\sigma$ (Meson-NH minus Observations), at Cerro Paranal, for the 129 nights sample (cf. Figure~\ref{fig:bias_rmse_all}).
The relative RMSE is also reported in the case of the wind direction.
The computations have been made considering the whole sample of data points.
For the wind direction, data corresponding to a wind speed inferior to 3~m$\cdot$s$^{-1}$ have been discarded from the computations.\label{tab:br_129n_par}}
 \begin{tabular}{|c|c|c||c|c||c|c||c|c|}
 \multicolumn{1}{c}{ } & \multicolumn{2}{c}{Temperature ($^{o}$C)} & \multicolumn{2}{c}{Wind speed (m$\cdot$s$^{-1}$)} & \multicolumn{2}{c}{Wind speed (m$\cdot$s$^{-1}$)} & \multicolumn{2}{c}{Wind direction ($^{o}$)} \\
 \multicolumn{1}{c}{ } & \multicolumn{2}{c}{$\Delta$X~=~500 m    } & \multicolumn{2}{c}{$\Delta$X~=~500 m            } & \multicolumn{2}{c}{$\Delta$X~=~100 m            } & \multicolumn{2}{c}{$\Delta$X~=~500 m      } \\
 \cline{2-9}
 \multicolumn{1}{c}{CERRO} & \multicolumn{2}{|c||}{ } & \multicolumn{2}{|c|}{ } & \multicolumn{2}{|c|}{ } & \multicolumn{2}{|c|}{ }\\
 \multicolumn{1}{c}{PARANAL} & \multicolumn{1}{|c}{2 m} & \multicolumn{1}{c||}{30 m} & \multicolumn{1}{|c}{10 m} & \multicolumn{1}{c|}{30 m} & \multicolumn{1}{|c}{10 m} & \multicolumn{1}{c|}{30 m} & \multicolumn{1}{|c}{10 m} & \multicolumn{1}{c|}{30 m} \\
\hline
 BIAS     &-0.12  & -0.37  &-2.90  &-1.39 &-1.16 &-0.24  &3.4    &-0.2\\
 RMSE     & 0.99  & 1.02   & 4.03  &3.04  &2.73 &2.70    &36.7   &35.7\\
 $\sigma$ & 0.98  & 0.95   & 2.80  & 2.70 &2.47 &2.69    &36.5   &35.7\\
 RMSE$_{rel}$ &   &        &       &      &     &        &20.4\% &19.8\%                           \\
 \hline
 \end{tabular}
\end{table*}
\begin{table*}
 \centering
 \caption{Individual nights model performance. Near surface median bias, RMSE and bias-corrected RMSE (Meson-NH minus Observations).
In small fonts, the 1st and 3rd quartiles. The values for the wind speed were obtained with the $\Delta$X~=~100~m configuration for Meso-NH, 
whereas the values obtained for the temperature and the wind direction were obtained with $\Delta$X~=~500~m configuration. \label{tab:br_ws_cumdist}} 
 \begin{tabular}{|c|c|c||c|c|c|c|}
 \cline{2-7}
 \multicolumn{1}{c}{ } & \multicolumn{2}{|c||}{Temperature} & \multicolumn{2}{|c|}{Wind speed} & \multicolumn{2}{|c|}{Wind direction} \\
 \multicolumn{1}{c}{ } & \multicolumn{1}{|c}{2 m} & \multicolumn{1}{c||}{30 m} &
 \multicolumn{1}{|c}{10 m} & \multicolumn{1}{c}{30 m} & \multicolumn{1}{c}{10 m} &
 \multicolumn{1}{c|}{30 m} \\
\hline
     & & & & & & \\
 BIAS  &     $-0.18_{-0.73}^{+0.40}$  & $-0.48_{-0.92}^{+0.12}$  &  $-0.85_{-2.50}^{+0.35}$  & $-0.14_{-1.86}^{+0.91}$   & $+3.90_{-12.22}^{+17.84}$  &  $-0.32_{-15.15}^{+12.06}$ \\
     & & & & & & \\
 RMSE  &     $0.91_{+0.60}^{+1.19}$   & $0.92_{+0.58}^{+1.25}$   &  $2.06_{+1.41}^{+3.09}$  & $2.30_{+1.63}^{+3.09}$   & $29.89_{+17.20}^{+43.94}$  & $27.29_{+14.57}^{+42.75}$  \\
     & & & & & & \\
 $\sigma$    & $0.54_{+0.38}^{+0.76}$   & $0.48_{+0.34}^{+0.69}$   &  $1.25_{+0.96}^{+1.70}$   & $1.45_{+1.09}^{+2.01}$  & $15.94_{+9.44}^{+27.32}$  & $15.30_{+9.06}^{+31.22}$  \\
     & & & & & & \\
 \hline
 \end{tabular}
\end{table*}
\begin{table*}
 \centering
 \caption{Winter: near surface median bias, RMSE and bias-corrected RMSE (Meson-NH minus Observations). Only the 54 winter nights are considered.
In small fonts, the 1st and 3rd quartiles. The values for the wind speed were obtained with the $\Delta$X~=~100~m configuration for Meso-NH,
whereas the values obtained for the temperature and the wind direction were obtained with $\Delta$X~=~500~m configuration. \label{tab:br_ws_cumdist_winter}}
 \begin{tabular}{|c|c|c||c|c|c|c|}
 \cline{2-7}
 \multicolumn{1}{c}{ } & \multicolumn{2}{|c||}{Temperature} & \multicolumn{2}{|c|}{Wind speed} & \multicolumn{2}{|c|}{Wind direction} \\
 \multicolumn{1}{c}{ } & \multicolumn{1}{|c}{2 m} & \multicolumn{1}{c||}{30 m} &
 \multicolumn{1}{|c}{10 m} & \multicolumn{1}{c}{30 m} & \multicolumn{1}{c}{10 m} &
 \multicolumn{1}{c|}{30 m} \\
\hline
     & & & & & & \\
 BIAS  &     $-0.33_{-0.81}^{+0.25}$  & $-0.53_{-1.04}^{+0.03}$  &  $-1.37_{-2.69}^{+0.34}$  & $+0.00_{-1.33}^{+1.02}$   & $+9.36_{-0.52}^{+18.35}$  &  $+5.02_{-5.08}^{+13.01}$ \\
     & & & & & & \\
 RMSE  &     $0.79_{+0.54}^{+1.16}$   & $0.80_{+0.52}^{+1.26}$   &  $2.21_{+1.34}^{+3.28}$  & $2.26_{+1.44}^{+3.45}$   & $18.65_{+14.61}^{+39.83}$  & $16.22_{+11.78}^{+33.09}$  \\
     & & & & & & \\
 $\sigma$    & $0.41_{+0.32}^{+0.60}$   & $0.39_{+0.30}^{+0.53}$   &  $1.08_{+0.82}^{+1.67}$   & $1.32_{+1.03}^{+1.95}$  & $11.12_{+7.97}^{+16.32}$  & $10.11_{+7.46}^{+17.41}$  \\
     & & & & & & \\
 \hline
 \end{tabular}
\end{table*}
\begin{table*}
 \centering
 \caption{Summer: near surface median bias, RMSE and bias-corrected RMSE (Meson-NH minus Observations). Only the 75 summer nights are considered.
In small fonts, the 1st and 3rd quartiles. The values for the wind speed were obtained with the $\Delta$X~=~100~m configuration for Meso-NH,
whereas the values obtained for the temperature and the wind direction were obtained with $\Delta$X~=~500~m configuration. \label{tab:br_ws_cumdist_summer}}
 \begin{tabular}{|c|c|c||c|c|c|c|}
 \cline{2-7}
 \multicolumn{1}{c}{ } & \multicolumn{2}{|c||}{Temperature} & \multicolumn{2}{|c|}{Wind speed} & \multicolumn{2}{|c|}{Wind direction} \\
 \multicolumn{1}{c}{ } & \multicolumn{1}{|c}{2 m} & \multicolumn{1}{c||}{30 m} &
 \multicolumn{1}{|c}{10 m} & \multicolumn{1}{c}{30 m} & \multicolumn{1}{c}{10 m} &
 \multicolumn{1}{c|}{30 m} \\
\hline
     & & & & & & \\
 BIAS  &     $-0.10_{-0.54}^{+0.53}$  & $-0.34_{-0.88}^{+0.16}$  &  $-0.81_{-1.96}^{+0.39}$  & $-0.31_{-2.06}^{+0.71}$   & $-2.51_{-17.37}^{+14.58}$  &  $-6.74_{-17.43}^{+10.51}$ \\
     & & & & & & \\
 RMSE  &     $0.94_{+0.66}^{+1.20}$   & $0.92_{+0.60}^{+1.25}$   &  $1.91_{+1.47}^{+2.88}$  & $2.42_{+1.78}^{+2.99}$   & $35.15_{+20.63}^{+48.89}$  & $32.65_{+18.56}^{+51.23}$  \\
     & & & & & & \\
 $\sigma$    & $0.62_{+0.46}^{+0.85}$   & $0.59_{+0.43}^{+0.73}$   &  $1.31_{+1.03}^{+1.71}$   & $1.54_{+1.18}^{+2.05}$  & $19.07_{+13.72}^{+37.22}$  & $20.66_{+13.66}^{+40.00}$  \\
     & & & & & & \\
 \hline
 \end{tabular}
\end{table*}

The forecast of the absolute temperature and the wind direction near the surface at Cerro Paranal, with the Meso-NH model, are excellent. Considering 
the whole sample (Figure~\ref{fig:bias_rmse_all} and Table~\ref{tab:br_129n_par}), the bias in the temperature is very small (not larger than 0.37$^{\circ}$C in absolute 
value) and the 
RMSE is not larger than 1$^{\circ}$C.

The bias in the wind direction is almost zero, and the RMSE is around 37$^{\circ}$ (the data for which the wind velocity was inferior to 3~m$\cdot$s$^{-1}$ 
have been excluded from the computations. 
These data are associated in general to a great dispersion for the wind direction but are not of interesting data for astronomers). 
This corresponds to a RMSE$_{rel}$ of around 20\%.

For the wind speed, we observe that the conclusions found in \cite{lascaux2013} are confirmed. The high horizontal resolution ($\Delta$X~=~100~m) is mandatory to reduce 
the bias to acceptable values (from 2.90~m$\cdot$s$^{-1}$ to 
1.16~m$\cdot$s$^{-1}$ at 10~m and from 1.39~m$\cdot$s$^{-1}$ to 0.24~m$\cdot$s$^{-1}$ at 30~m, in absolute values). 
With the $\Delta$X~=~100~m resolution, the RMSE is around 2.7~m$\cdot$s$^{-1}$.\\ \\
When looking at the single nights statistics, the performances are even better.
We computed the bias, RMSE and bias-corrected RMSE for every night of the sample, and constructed the cumulative distributions at every levels for every meteorological 
parameter.
From the cumulative distributions we can extract the median and the first and third quartiles. 
Figure~\ref{fig:cumdist} and \ref{fig:cumdist2} in the appendix show the cumulative distributions at all levels. 
Table~\ref{tab:br_ws_cumdist} summarizes all the median values, together with the 
first and third quartiles, for the single nights bias, RMSE and $\sigma$, for the absolute temperature, the wind speed with the $\Delta$X~=~100~m configuration, and the wind 
direction. 
The median RMSE for the absolute temperature is less than 0.92$^{\circ}$C. 
It is inferior to 2.30~m$\cdot$s$^{-1}$ for the wind speed, and inferior to 29$^{\circ}$ for the wind direction.\\ 

To study the seasonal variation and check if some systematic better/worst model performance is observed in summer and winter we also calculated bias, RMSE and 
bias-corrected RMSE for the two periods of [April-September] that we call winter and [October-March] that we call summer 
(Table \ref{tab:br_ws_cumdist_winter} and Table \ref{tab:br_ws_cumdist_summer}).
The two periods are made of 54 and 75 nights, respectively. Summer period has more nights because of the presence of the first 20 nights analyzed by \cite{lascaux2013} that were all concentrated in the months of November and December 2007. 
We observe that, for the temperature, in summer we have a slightly better median bias but a slightly larger median RMSE as well as sigma (bias-corrected RMSE) than in winter. In any case, the median RMSE remains below 1$^{\circ}$C.
For the wind direction the median bias of the absolute temperature is sightly better in summer than in winter but the median RMSE is larger (almost the double 33-35$^{\circ}$ instead of 16-19$^{\circ}$) in summer than in winter. The sigma also shows the same trend even if the difference between summer and winter is less marked. For the wind speed negligible difference are observed for the median bias and RMSE.\\

Now that we have characterized the performances of the model using standards statistical parameters (bias, RMSE and $\sigma$), we want to know how good or bad 
is the model in predicting some specific events (like strong wind speed) or categories (like the quadrants from which the wind is blowing).
For this we have constructed contingency tables.
\clearpage
\section{What is a contingency table?}
\label{sec4}
\begin{table*}
\begin{center}
\caption{\label{tab:ct_gen22} Generic 2$\times$2 contingency table.}
\begin{tabular}{ccccc}
\multicolumn{2}{c}{\multirow{2}{*}{EVENT}} & \multicolumn{2}{c}{\bf OBSERVATIONS} &\\
\multicolumn{2}{c}{ } & YES & NO & Total \\
\hline
\multirow{7}{*}{\rotatebox{90}{\bf MODEL}} & & &  &\\
 & \multirow{2}{*}{YES}  & a  &  b  &   a+b \\
 &           & (hit) & (false alarm) &  Yes (Model)\\
 &      &    &    &    \\
 & \multirow{2}{*}{NO}  & c  & d & c+d\\
 &           &   (miss) & (correct rejection) & No (Model)\\
 &      &    &   &     \\
\hline
 & \multirow{2}{*}{Total} & a+c             &  b+d          & N=a+b+c+d \\
 &                                       & Yes (OBS) & No (OBS) & Total of events \\
\end{tabular}
\end{center}
\end{table*}
\begin{table*}
\begin{center}
\caption{\label{tab:ct_gen33} Generic 3$\times$3 contingency table.}
\begin{tabular}{cccccc}
\multicolumn{2}{c}{\multirow{2}{*}{Intervals}} & \multicolumn{3}{c}{\bf OBSERVATIONS} &\\
\multicolumn{2}{c}{ } & 1 & 2 & 3 & Total \\
\hline
\multirow{10}{*}{\rotatebox{90}{\bf MODEL}} & & &  &\\
 & \multirow{2}{*}{1}  & a  &  \multirow{2}{*}{b}  &  \multirow{2}{*}{c} &  a+b+c \\
 &           & (hit 1) & & &  1 (Model)\\
 &      &    &    &  &  \\
 & \multirow{2}{*}{2}  & \multirow{2}{*}{d}  & e & \multirow{2}{*}{f} & d+e+f\\
 &           &    & (hit 2) & & 2 (Model)\\
 &      &    &   &  &   \\
 & \multirow{2}{*}{3}  & \multirow{2}{*}{g}  & \multirow{2}{*}{h} & i & g+h+i\\
 &           &    &  & (hit 3) & 3 (Model)\\
 &      &    &   &  &   \\
\hline
 & \multirow{2}{*}{Total} & a+d+g           &  b+e+h        & c+f+i & N=a+b+c+d+e+f+g+h+i \\
 &                        & 1 (OBS) & 2 (OBS) & 3 (OBS) & Total of events \\
\end{tabular}
\end{center}
\end{table*}
\begin{table*}
\begin{center}
\caption{\label{tab:ct_gen44} Generic 4$\times$4 contingency table.}
\begin{tabular}{ccccccc}
\multicolumn{2}{c}{\multirow{2}{*}{Intervals}} & \multicolumn{3}{c}{\bf OBSERVATIONS} &\\
\multicolumn{2}{c}{ } & 1 & 2 & 3 & 4 & Total \\
\hline
\multirow{13}{*}{\rotatebox{90}{\bf MODEL}} & & &  &\\
 & \multirow{2}{*}{1}  & a  &  \multirow{2}{*}{b}  &  \multirow{2}{*}{c} &  \multirow{2}{*}{d} & a+b+c+d \\
 &           & (hit 1) & & & &  1 (Model)\\
 &      &    &    &  &  & \\
 & \multirow{2}{*}{2}  & \multirow{2}{*}{e}  & f & \multirow{2}{*}{g} & \multirow{2}{*}{h} & e+f+g+h\\
 &           &    & (hit 2) & & & 2 (Model)\\
 &      &    &   &  &   \\
 & \multirow{2}{*}{3}  & \multirow{2}{*}{i}  & \multirow{2}{*}{j} & k & \multirow{2}{*}{l} & i+j+k+l\\
 &           &    &  & (hit 3) & & 3 (Model)\\
 &      &    &   &  &   \\
 & \multirow{2}{*}{4}  & \multirow{2}{*}{m}  & \multirow{2}{*}{n} & \multirow{2}{*}{o} & p & m+n+o+p\\
 &           &    &  & & (hit 4) & 4 (Model)\\
 &      &    &   &  &   \\
\hline
 & \multirow{2}{*}{Total} & \multirow{2}{*}{a+e+i+m} &  \multirow{2}{*}{b+f+j+n} & \multirow{2}{*}{c+g+k+o} & \multirow{2}{*}{d+h+l+p} & N=a+b+c+d+e+f+g+h   \\
 &                        &                          &                           &                          & \                        &  +i+j+k+l+m+n+o+p     \\
 &                        & 1 (OBS) & 2 (OBS) & 3 (OBS) & 4 (OBS) & Total of events \\
\end{tabular}
\end{center}
\end{table*}
Like mentioned in the Introduction, a contingency table allows for the analysis of the relationship between two or more categorical variables.
Table~\ref{tab:ct_gen22} is an example of a generic 2$\times$2 contingency table, where the observations are divided in 2 categories (YES and NO), 
that could for example correspond to a wind above a given threshold (YES) and a wind below this threshold (NO). 
In the 2$\times$2 contingency table, four combinations are possible:\\
-'$a$' is the number of times an event was well reproduced by the model (hit);\\
-'$b$' is the number of times an event was erroneously forecasted by the model (false alarms);\\ 
-'$c$' is the number of times an event was not reproduced by the model (miss);\\
-'$d$' is the number of times the absence of an event was well reproduced by the model (correct rejection).\\
Using $a$, $b$, $c$ and $d$ (and $N=a+b+c+d$), 
we can compute different probabilities useful to have an insight on how well (or bad) the model performed for a particular event.
\par
Hereafter we list all the different simple scores we will use in the following of the article 
(to define them we use the numbers $a$, $b$, $c$, $d$ and $N$ from the generic 2$\times$2 contingency table of Table~\ref{tab:ct_gen22}):\\
-the percent of correct detections $PC$ (in \%):
\begin{equation}
PC=\frac{a+d}{N}\times 100; 0\%\leq PC\leq 100\%
\label{eq:pc}
\end{equation}
$PC=$100\% is the best score, and corresponds to a perfect forecast. 
The $PC$ alone is not sufficient because sometimes it can be influenced by the most common category and it 
does not provide us informations on the ability of the model in detecting observations in the different categories. 
It is therefore always preferable to provide also the probability of detection ($POD$) together to the $PC$.\\
-the probability of detection ($POD$, in \%) of a given event, or hit-rate:
\begin{equation}
POD(event_1)=\frac{a}{a+c} \times 100; 0\%\leq POD\leq 100\%
\label{eq:pod}
\end{equation}
$POD=$100\% is the best score. $POD$ measures the proportion of observed events that have been correctly predicted by the model. 
In the 2$\times$2 contingency table, only one event is generally considered, and thus only one $POD$ is calculated.
However we could also consider the non occurrence of the event as an event itself, and then define a second $POD$ as:
\begin{equation}
POD(event_2)=\frac{d}{b+d} \times 100; 0\%\leq POD\leq 100\%
\label{eq:pod_2}
\end{equation}
It can be of interest to highlight that, for a total random prediction and in the case of a 2$\times$2 contingency table,
$a=b=c=d$.
That means that all $POD$ are equal to 50\%, and $PC$=50\%.
The model is useful if it performs better than this random case.\\
A huge number of others statistical parameters can be deduced from a contingency table (such as false alarm ratio, probability of false detection, etc.). 
We will limit the current study to the analysis of the parameters mentioned above that already provide a consistent panorama of the model potentialities.\\
For our purpose a 2$\times$2 table is, however, a too simplified analysis. 
A 3$\times$3 table is definitely more appropriated. 
It consists in dividing the observed values in three categories delimited by some thresholds. 
An example of a 3$\times$3 contingency tables is shown in Table~\ref{tab:ct_gen33}. 
Equivalent $PC$ and $POD$ can be defined using $a$, $b$, $c$, $d$, $e$, $f$, $g$, $h$, $i$ and $N$ from Table~\ref{tab:ct_gen33}. 
In the case of the 3$\times$3 contingency tables, we also add a third parameter we call $EBD$ (for extremely bad detection).
\begin{equation}
PC=\frac{a+e+i}{N} \times 100
\label{eq:pc2}
\end{equation}
\begin{equation}
POD(event_1)=\frac{a}{a+d+g} \times 100
\label{eq:pod1}
\end{equation}
\begin{equation}
POD(event_2)=\frac{e}{b+e+h} \times 100
\label{eq:pod2}
\end{equation}
\begin{equation}
POD(event_3)=\frac{i}{c+f+i} \times 100
\label{eq:pod3}
\end{equation}
\begin{equation}
EBD=\frac{c+g}{N} \times 100
\label{eq:ebd}
\end{equation}
EBD represents the percent of the most distant predictions, by the model, from the observations.
For a perfect forecast, it is equal to 0\%.
In the case of a perfectly random forecast ($a=b=...=i=\frac{N}{9}$), all $POD$ are equal to 33\%, $PC$=33\%, and $EBD$=22.2\%.
These values are a good reference to evaluate the performances of the model. \\
For the wind direction a 3$\times$3 table is not appropriated because this is a circular variable. A 4$\times$4 (divided in four quadrants with 90$^{\circ}$ each) is therefore a more suitable solution (see Table~\ref{tab:ct_gen44}). $PC$ and $POD$ are defined in this case as:
\begin{equation}
PC=100\times\frac{a+f+k+p}{N}
\label{eq:pc3}
\end{equation}
\begin{equation}
POD(event_1)=100\times\frac{a}{a+e+i+m}
\label{eq:pod1b}
\end{equation}
\begin{equation}
POD(event_2)=100\times\frac{f}{b+f+j+n}
\label{eq:pod2b}
\end{equation}
\begin{equation}
POD(event_3)=100\times\frac{k}{c+g+k+o}
\label{eq:pod3b}
\end{equation}
\begin{equation}
POD(event_4)=100\times\frac{p}{d+h+l+p}
\label{eq:pod4b}
\end{equation}
\begin{equation}
EBD=100\times\frac{c+h+i+n}{N}
\label{eq:ebd2}
\end{equation}
If we consider the random case ($a=b=...=p=\frac{N}{16}$), $PC$=25\%, all $POD$ are equal to 25\%, and $EBD$=25\%.\\
From this point and until the end of the article, we will write $POD_i$ instead of $POD(event_i)$ with $i$ the event considered.
\section{Climatological tertiles at Cerro Paranal}
\label{sec:climato}
To construct the contingency tables that we will discuss in the next sections, it is necessary to define some thresholds limiting the discretized intervals of the observed values and to divide 
the sample in three categories.
In the cases of the 3$\times$3 contingency tables, we need two thresholds.
In this study, we decided to use the climatological tertiles computed thanks to the available measurements.
We used 6 complete years of measurements from a mast, from January 1, 2006 to December 31, 2012, that are part
of the VLT Astronomical Site Monitor \citep{sandrock99}.
At each level, and for each parameter, we computed the median value, and the tertiles, of all the available measurements.
Table~\ref{tab:climato_ws_par} shows the results for the wind speed at Cerro Paranal.
Tables~\ref{tab:climato_t_par} shows the results for the absolute temperature at Cerro Paranal.
\begin{table}
\begin{center}
\caption{\label{tab:climato_ws_par} Climatological tertiles for the wind speed at Cerro Paranal (left column: 33\%, central column: median, right column: 66\%)}
\begin{tabular}{cccc}
\hline
Cerro  & \multicolumn{3}{c}{Wind Speed (in $m\ s^{-1}$)} \\
Paranal & 33\% & Median (50\%) & 66\% \\
& & & \\
10 m & 3.98 & 5.47 & 7.25 \\
30 m & 4.40 & 6.10 &  8.10 \\
\hline
\end{tabular}
\end{center}
\end{table}
\begin{table}
\begin{center}
\caption{\label{tab:climato_t_par} Climatological tertiles for the absolute temperature 
at Cerro Paranal (left column: 33\%, central column: median, right column: 66\%)}
\begin{tabular}{cccc}
\hline
Cerro  & \multicolumn{3}{c}{Temperature (in Celsius)} \\
Paranal & 33\% & Median (50\%) & 66\% \\
& & & \\
2 m &11.03 &12.24 &13.27 \\
30 m &11.40 &12.58 & 13.64 \\
\hline
\end{tabular}
\end{center}
\end{table}
%
\section{Absolute temperature}
\label{sec6}
The first meteorological parameter reconstructed by the model that we investigate is the absolute temperature near the ground.
We report the results only for the standard model ($\Delta$X=500~m). 
We highlight that results obtained with the high horizontal resolution model ($\Delta$X=100~m) (not shown here) are very similar. 
Table~\ref{tab:ct_cp2_temp} and Table~\ref{tab:ct_cp30_temp} are the contingency tables for the absolute temperature 
at Cerro Paranal, at 2~m and 30~m, respectively. Each estimate corresponds to the average over a 20-minutes interval (as indicated
in Section \ref{sec3}).
For sake of simplification, we used for thresholds, the rounded values taken from the climatoligical study of Section~\ref{sec:climato}.
We can observe that the percent of good forecast by the model (characterized by $PC$) is very good at every level.
$PC$ is between 73\% and 75\%.
More over, an $EBD$ very close to 0\% at both levels, is the sign that the model never produces extremely bad forecasts 
for the temperature.
If we look at the probability of detection of the temperature in a given interval (characterized by $POD$), once again the model demonstrates 
its accuracy.
$POD$ is between 55.2\% (detection of temperature between 11.5$^{\circ}$C and 
13.5$^{\circ}$C at 30~m) and 93.7\% 
(detection of a temperature inferior to 11.5$^{\circ}$C at 30~m). In all cases, PC and PODs are well larger than 33$\%$ (value of random case). 
This proves therefore the utility of the model.
\begin{table*}
\begin{center}
\caption{\label{tab:ct_cp2_temp} 3$\times$3 contingency table for the absolute temperature during the night, at 2~m a.g.l. at Cerro Paranal, for the sample of 129 nights.
We use the Meso-NH ${\Delta}$X~=~500~m configuration.}
\begin{tabular}{ccccc}
\multicolumn{2}{c}{Division by tertiles (climatology)} & \multicolumn{3}{c}{\bf OBSERVATIONS}\\
\multicolumn{2}{c}{C. Paranal - 2~m} & $T<11^{\circ}C$ & $11^{\circ}C<T<13.5^{\circ}C$   &  $T>13.5^{\circ}C$ \\
\hline
\multirow{7}{*}{\rotatebox{90}{\bf MODEL}} & & &\\
 & $T<11^{\circ}C$               & 1157 & 315 & 0\\
 &      &  &  & \\
 & $11^{\circ}C<T<13.5^{\circ}C$ & 96 & 972 & 258\\
 &      &  &  & \\
 & $T>13.5^{\circ}C$             & 8 & 178 & 499\\
 &      &  &  & \\
\hline
\multicolumn{5}{l}{Total points = 3483; $PC$=75.5\%; $EBD$=0.2\%} \\
\multicolumn{5}{l}{$POD_1$=91.8\%; $POD_2$=66.3\%; $POD_3$=65.9\%} \\
\end{tabular}
\end{center}
\end{table*}
\begin{table*}
\begin{center}
\caption{\label{tab:ct_cp30_temp} 3$\times$3 contingency table for the absolute temperature during the night, at 30~m a.g.l. at Cerro Paranal, for the sample of 129 nights.
We use the Meso-NH ${\Delta}$X~=~500~m configuration.}
\begin{tabular}{ccccc}
\multicolumn{2}{c}{Division by tertiles (climatology)} & \multicolumn{3}{c}{\bf OBSERVATIONS}\\
\multicolumn{2}{c}{C. Paranal - 30~m} & $T<11.5^{\circ}C$ & $11.5^{\circ}C<T<13.5^{\circ}C$   &  $T>13.5^{\circ}C$ \\
\hline
\multirow{7}{*}{\rotatebox{90}{\bf MODEL}} & & &\\
 & $T<11.5^{\circ}C$               & 1250 & 422 & 2\\
 &      &  &  & \\
 & $11.5^{\circ}C<T<13.5^{\circ}C$ & 68 & 663 & 292\\
 &      &  &  & \\
 & $T>13.5^{\circ}C$               & 16 & 117 & 653\\
 &      &  &  & \\
\hline
\multicolumn{5}{l}{Total points = 3483; $PC$=73.7\%; $EBD$=0.5\%} \\
\multicolumn{5}{l}{$POD_1$=93.7\%; $POD_2$=55.2\%; $POD_3$=69.0\%} \\
\end{tabular}
\end{center}
\end{table*}
%
\section{Wind speed}
\label{sec7}
Considering the tendency of the model to slightly underestimate the wind velocity (see section~\ref{sec3}), 
(even if this underestimation is obviously quantitatively different between $\Delta$X=500~m and $\Delta$X=100~m), 
it was decided to evaluate the predictive power of the model using a forecasted wind speed corrected by the multiplicative bias $B_{mult}$ computed as:
\begin{equation}
B_{mult}(k)=\frac{\overline{WS_{mod}}(k)}{\overline{WS_{obs}}(k)}
\label{eq:b_mult}
\end{equation}
with $\overline{WS_{mod}}(k)$ the average of the forecasted wind speed over the whole sample of the 129 nights, and
$\overline{WS_{obs}}(k)$ the average of the observed wind speed from the same sample, at the level k (10~m and 30~m for
Cerro Paranal). The different values are summarized in Table~\ref{tab:b_mult}. 
The choice of the multiplicative bias is justified by the fact that we want the highest values of the wind speed to be more corrected than the lowest values.
It is important to note that the values of $B_{mult}(k)$ can be considered as constant, whatever the sub-sample considered. 
Indeed, we have computed it with the initial sample of 20 nights, with another subsample of 73 nights, and with the final sample of 129 nights, and in all cases its values 
were the one reported in Table~\ref{tab:b_mult} (variations of the second digit only are noticeable, and thus not reported here). This tells us that is useless to consider a calibration sub-sample and a complementary and independent validation or testing sub-sample. It is therefore more appropriate to apply this multiplicative factor on the whole sample night.
\begin{table}
\begin{center}
\caption{\label{tab:b_mult} Wind speed multiplicative biases (from Eq.~\ref{eq:b_mult}). Values are rounded.}
\begin{tabular}{|c|c|c|c|c|}
\multicolumn{1}{c|}{ } & 10 m & 30 m \\
\hline
C. Paranal        & \multirow{2}{*}{0.5} & \multirow{2}{*}{0.8} \\
${\Delta}X$ = 500 m & & & &\\
\hline
C. Paranal        & \multirow{2}{*}{0.8} & \multirow{2}{*}{0.9} \\
${\Delta}X$ = 100 m & & & & \\
\hline
\end{tabular}
\end{center}
\end{table}
\subsection{Sensibility of the model detection to the wind speed threshold}
\label{sec7_sub1}
\begin{figure}
\begin{center}
\begin{tabular}{c}
\includegraphics{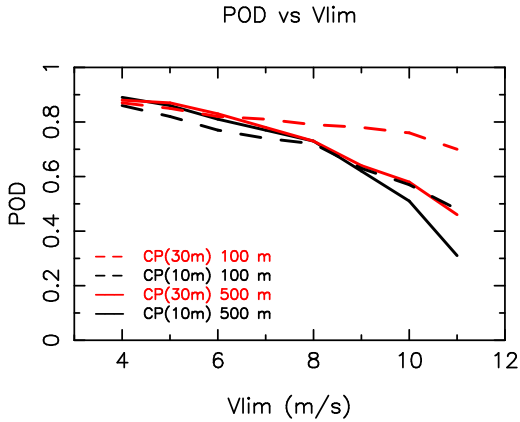}
\end{tabular}
\end{center}
\caption[vlim]{\label{fig:vlim} Evolution of the probability of detection ($POD$) from 2$\times$2 contingency tables, 
by Meso-NH, of winds stronger than $V_{lim}$, in function of $V_{lim}$.
Dashed lines, with ${\Delta}X$~=~100~m configuration.
Continuous lines, with ${\Delta}X$~=~500~m configuration.
In all cases, the Meso-NH wind speed was corrected by the multiplicative bias of Table~\ref{tab:b_mult}.
CP stands for Cerro Paranal.}
\end{figure}
Considering that the most critical issue for the wind speed is the model performances in detecting 
the strong wind speed we performed the following exercise before to calculate the 3$\times$3 contingency tables. 
To see how the model outputs depend on the choice of the threshold used for the wind speed, we plotted on Fig.~\ref{fig:vlim} the dependence of 
$POD_1$ from 2$\times$2 contingency tables, on the threshold $V_{lim}$, for both ${\Delta}X$~=~500~m and ${\Delta}X$~=~100~m configurations.
$V_{lim}$ varies between 4~$m{\cdot}s^{-1}$ and 11~$m{\cdot}s^{-1}$. 
\begin{table}
\begin{center}
\caption{\label{tab:ct_vlim_6_cp10_100m_bmult} Contingency table considering the event: winds stronger than 6~$m{\cdot}s^{-1}$. The site
is Cerro Paranal. The altitude is 10~m a.~g.~l. The wind speed is computed with the Meso-NH ${\Delta}X$~=~100~m configuration and corrected by
the multiplicative bias from Table~\ref{tab:b_mult}.}
\begin{tabular}{cccc}
\multicolumn{2}{c}{$WS > 6~m{\cdot}s^{-1}$} & \multicolumn{2}{c}{\bf OBSERVATIONS}\\
\multicolumn{2}{c}{C. Paranal - 10m} & YES & NO  \\
\hline
\multirow{5}{*}{\rotatebox{90}{\bf MODEL}} & & &\\
 & YES  & 1230 & 418 \\
 &      &  &  \\
 & NO   & 363 & 1445 \\
 &      &  &  \\
\hline
\multicolumn{4}{l}{Total points = 3456} \\
\multicolumn{4}{l}{$PC$=77.4\%} \\
\multicolumn{4}{l}{$POD_1$(WS$>$6$m{\cdot}s^{-1}$)=77.2\%} \\ 
\multicolumn{4}{l}{$POD_2$(WS$<$6$m{\cdot}s^{-1}$)=77.6\%} \\
\end{tabular}
\end{center}
\end{table}
The two Meso-NH configurations ($\Delta$X = 500 and 100 meters) were used with the correction of the wind by the multiplicative bias.
The first thing we can notice in Fig.\ref{fig:vlim} is that, the performance of the simulations with
the ${\Delta}X$~=~100~m configuration depends less on the threshold than the performance of the
${\Delta}X$~=~500~m configuration. 
Up to $V_{lim}$~=~10~$m{\cdot}s^{-1}$, with ${\Delta}X$~=~100~m all $POD_1$ are very good with 
the worst PODs at 10~$m{\cdot}s^{-1}$ between 70\% and 80\%.
On the contrary, the model performance decreases faster with $V_{lim}$, for the ${\Delta}X$~=~500~m configuration.
At $V_{lim}$~=~9~$m{\cdot}s^{-1}$, the performance is still satisfactory (all $POD_1$ are around 60\%), but for $V_{lim}>$ 9 $m{\cdot}s^{-1}$, the
performances are degraded (all $POD_1$ are inferior to 50\%, down to $\sim$30\% for $V_{lim}$~=~11~$m{\cdot}s^{-1}$). 
This tells us that the 100~m configuration guarantees a better model performances than the 500~m one.
As an example, if we consider at Cerro Paranal a reasonable value of $V_{lim}$=6~$m{\cdot}s^{-1}$ and we study the probability to detect wind speed stronger 
than this threshold we obtain very satisfactory results (Table~\ref{tab:ct_vlim_6_cp10_100m_bmult}) for the PODs that are typically of the order of 77$\%$.\\ 

The interest for astronomers is mainly to be able to identify conditions of strong wind and this result tells us that such a model in this configuration well answers to this necessity.

\subsection{Wind speed 3$\times$3 contingency tables}
\label{sec7_sub2}
In this section we investigate the ability of the model in forecasting the wind speed as we did in Section \ref{sec6}.
As we have seen in Section \ref{sec7_sub1}, we consider the best configuration for the model i.e. the high horizontal ($\Delta$X=100~m) resolution configuration.
The contingency tables have been divided in three intervals, for each level concerned, using the rounded climatological tertiles 
from section~\ref{sec:climato} as thresholds.
They are reported in Tables~\ref{tab:ct_cp10_ws} and \ref{tab:ct_cp30_ws}, at 10~m and 30~m, respectively. \\
$PC$ is significatively better than 33\% (random case): it is 60.0\% at 10~m and 60.6\% at 30~m.
The $POD$ also are very satisfactory, between 47.5\% and 79.0\%. More over, the $EBD$ are negligible (less than 4.4\%).
If we consider $POD_3$ only (that is, detection of the wind superior to a fixed threshold, the 
2$^{nd}$ climatological tertile), that is the most interesting informations for astronomers, the results are even better ($\sim$79.0\%).\\
A strong wind is indeed the main cause of vibrations of the primary mirror and adaptive secondary (a critical element of the adaptive optics system).
All $POD_3$  for both sites and at all altitudes are excellent, between 74.3\% and 79.0\%.
This demonstrates the ability of the model in predicting at Cerro Paranal critical wind speed values for AO applications.
\begin{table*}
\begin{center}
\caption{\label{tab:ct_cp10_ws} 3$\times$3 contingency table for the wind speed during the night, at 10~m a.g.l. at Cerro Paranal.
We use the Meso-NH ${\Delta}$X~=~100~m configuration with the wind corrected by the multiplicative bias.}
\begin{tabular}{ccccc}
\multicolumn{2}{c}{Division by tertiles (climatology)} & \multicolumn{3}{c}{\bf OBSERVATIONS}\\
\multicolumn{2}{c}{C. Paranal - 10~m} & $T<4~m{\cdot}s^{-1}$ & $4~m{\cdot}s^{-1}<T<7~m{\cdot}s^{-1}$   &  $T>7~m{\cdot}s^{-1}$ \\
\hline
\multirow{7}{*}{\rotatebox{90}{\bf MODEL}} & & &\\
 & $T<4~m{\cdot}s^{-1}$                     & 641 & 283&50\\
 &      &  &  & \\
 & $4~m{\cdot}s^{-1}<T<7~m{\cdot}s^{-1}$ & 395&510&269\\
 &      &  &  & \\
 & $T>7~m{\cdot}s^{-1}$                   &103&281 &924\\
 &      &  &  & \\
\hline
\multicolumn{5}{l}{Total points = 3456; $PC$=60.0\%; $EBD$=4.4\%} \\
\multicolumn{5}{l}{$POD_1$=56.2\%; $POD_2$=47.5\%; $POD_3$=74.3\%} \\
\end{tabular}
\end{center}
\end{table*}
\begin{table*}
\begin{center}
\caption{\label{tab:ct_cp30_ws} 3$\times$3 contingency table for the wind speed during the night, at 30~m a.g.l. at Cerro Paranal.
We use the Meso-NH ${\Delta}$X~=~100~m configuration with the wind corrected by the multiplicative bias.}
\begin{tabular}{ccccc}
\multicolumn{2}{c}{Division by tertiles (climatology)} & \multicolumn{3}{c}{\bf OBSERVATIONS}\\
\multicolumn{2}{c}{C. Paranal - 30~m} & $T<4~m{\cdot}s^{-1}$ & $4~m{\cdot}s^{-1}<T<8~m{\cdot}s^{-1}$   &  $T>8~m{\cdot}s^{-1}$ \\
\hline
\multirow{7}{*}{\rotatebox{90}{\bf MODEL}} & & &\\
 & $T<4~m{\cdot}s^{-1}$                     & 554 &289 &27\\
 &      &  &  & \\
 & $4~m{\cdot}s^{-1}<T<8~m{\cdot}s^{-1}$ & 400&559&234\\
 &      &  &  & \\
 & $T>8~m{\cdot}s^{-1}$                   &101& 311&981\\
 &      &  &  & \\
\hline
\multicolumn{5}{l}{Total points = 3456; $PC$=60.6\%; $EBD$=3.7\%} \\
\multicolumn{5}{l}{$POD_1$=52.5\%; $POD_2$=48.2\%; $POD_3$=79.0\%} \\
\end{tabular}
\end{center}
\end{table*}
%
\section{Wind direction}
\label{sec8}
\begin{figure*}
\begin{center}
\begin{tabular}{c}
\includegraphics[angle=270,width=0.8\textwidth]{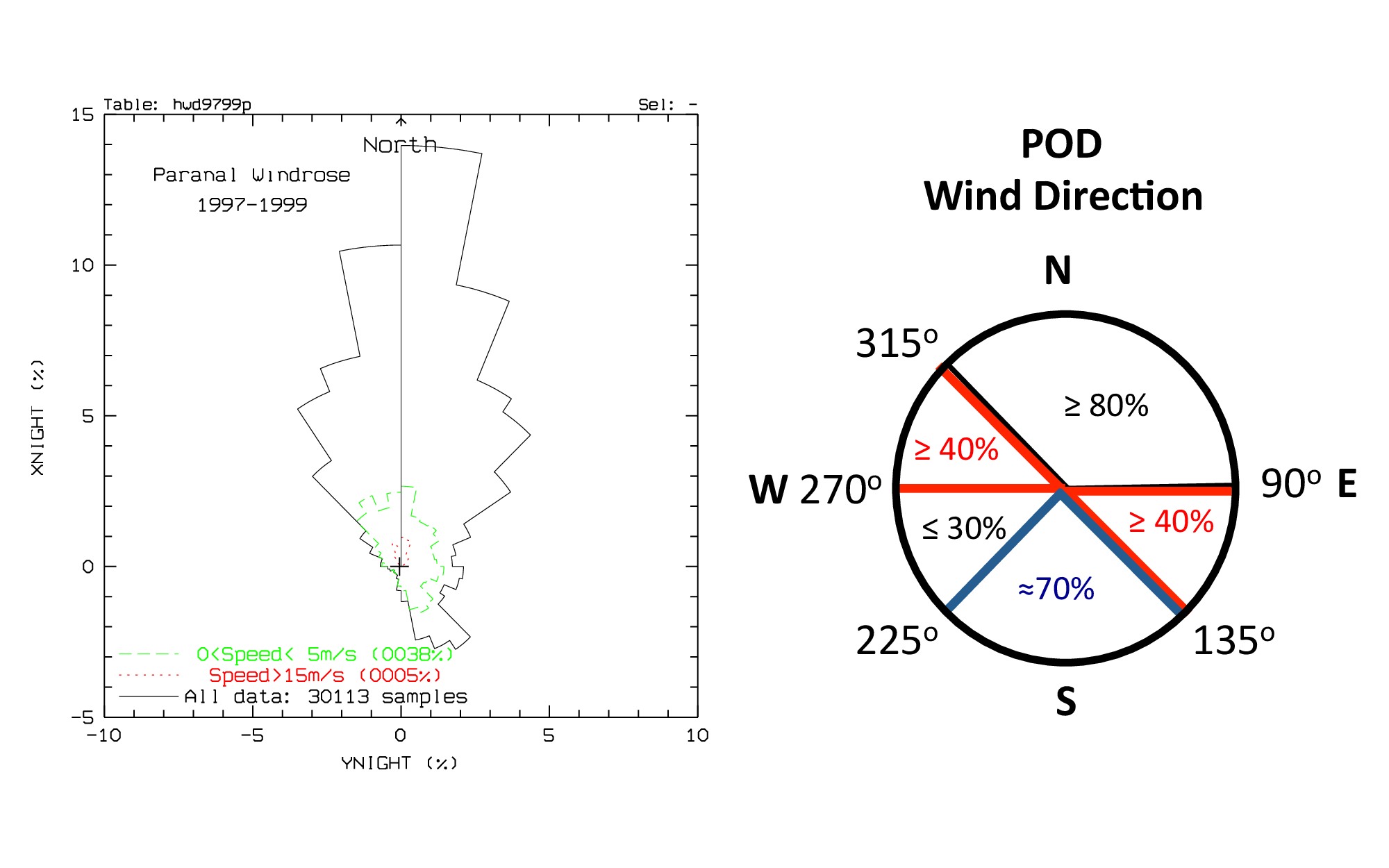}\\
\end{tabular}
\end{center}
\caption{\label{fig:windrose} Left panel: observed wind direction at 10~m above the ground at Cerro Paranal.
The sectors are in steps of 11.25 degrees, the length of the sector represents the fraction of time when the wind comes from this direction.
The figure is extracted from the ESO astroclimatic webpage at $https://www.eso.org/gen-fac/pubs/astclim/paranal/wind$. Right panel: the probability of detection
(POD) of the wind direction by the model is summarized by sectors. The circle has been constructed by combining the results from Tables~\ref{tab:ct_cp_wd}
and \ref{tab:ct_cp_wd2}.}
\end{figure*}

In this section we investigate the reconstruction of the wind direction by the Meso-NH model with the ${\Delta}$X~=~500~m configuration.
We have constructed 4$\times$4 contingency tables divided in four quadrants of 90$^{\circ}$ each.
\begin{figure}
\begin{center}
\begin{tabular}{c}
\includegraphics[width=0.30\textwidth]{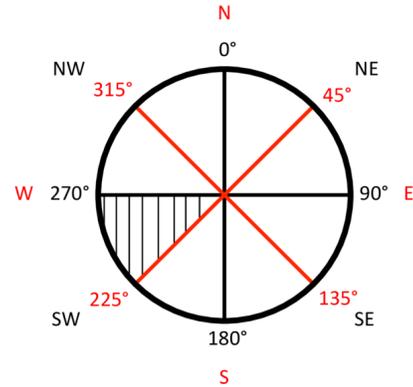}
\end{tabular}
\end{center}
\caption[wind_circle]{\label{fig:wind_circle} Convention chosen for the wind direction ($0^{\circ}<\alpha<360^{\circ}$). 
The black quadrants correspond to the 
intervals of 90$^{\circ}$ chosen 
for the 4$\times$4 contingency tables of Table~\ref{tab:ct_cp_wd}; 
the red quadrants correspond to the intervals of 90$^{\circ}$ chosen for the 
4$\times$4 contingency tables of Table~\ref{tab:ct_cp_wd2}. 
The hatched section is discussed in the text.}
\end{figure}
\clearpage
\begin{table}
\begin{center}
\caption{\label{tab:ct_cp_wd} 4$\times$4 contingency table for the wind direction $\alpha$ during the night, at 10~m, and 30~m a.g.l. at Cerro Paranal.
We use the Meso-NH ${\Delta}$X~=~500~m configuration. We filter out the observed wind inferior to 3~$m{\cdot}s^{-1}$.
NE corresponds to $0^{\circ}<{\alpha}<90^{\circ}$; SE corresponds to $90^{\circ}<{\alpha}<180^{\circ}$;
SW corresponds to $180^{\circ}<{\alpha}<270^{\circ}$; NW corresponds to $270^{\circ}<{\alpha}<360^{\circ}$.
$\alpha$ is the wind direction (see Fig.~\ref{fig:wind_circle}).}
\begin{tabular}{cccccc}
\multicolumn{2}{c}{ } & \multicolumn{4}{c}{\bf OBSERVATIONS}\\
\multicolumn{2}{c}{C. Paranal - all levels} & N-E & S-E & S-W & N-W \\
\hline
\multirow{9}{*}{\rotatebox{90}{\bf MODEL}} & & &\\
 & N-E  &2809& 212& 35 &815 \\
 &     & & & &\\
 & S-E  &102 & 734& 109& 28 \\
 &     & & & &\\
 & S-W  & 23 & 87 & 63 & 11 \\
 &     & & & &\\
 & N-W  & 505& 87 & 19 &614 \\
 &     & & & &\\
\hline
\multicolumn{5}{l}{Total points = 6253; $PC$=67.5\%; $EBD$=2.8\%} \\
\multicolumn{5}{l}{$POD(NE)$=81.7\%; $POD(SE)$=65.5\%} \\
\multicolumn{5}{l}{$POD(SW)$=27.9\%; $POD(NW)$=41.8\%} \\
\end{tabular}
\end{center}
\end{table}
\begin{table}
\begin{center}
\caption{\label{tab:ct_cp_wd2} 4$\times$4 contingency table for the wind direction $\alpha$ during the night, at 10~m, and 30~m a.g.l. at Cerro Paranal.
We use the Meso-NH ${\Delta}$X~=~500~m configuration. We filter out the observed wind inferior to 3~$m{\cdot}s^{-1}$.
N corresponds to $-45^{\circ}<{\alpha}<45^{\circ}$; E corresponds to $45^{\circ}<{\alpha}<135^{\circ}$;
S corresponds to $135^{\circ}<{\alpha}<225^{\circ}$; W corresponds to $225^{\circ}<{\alpha}<315^{\circ}$.
$\alpha$ is the wind direction (see Fig.~\ref{fig:wind_circle}).}
\begin{tabular}{cccccc}
\multicolumn{2}{c}{ } & \multicolumn{4}{c}{\bf OBSERVATIONS}\\
\multicolumn{2}{c}{C. Paranal - all levels} & N & E & S & W \\
\hline
\multirow{9}{*}{\rotatebox{90}{\bf MODEL}} & & &\\
 & N  &3807&536 & 68 &138 \\
 &     & & & &\\
 & E  &196 &473&119 & 4  \\
 &     & & & &\\
 & S  &18  &141 & 512  & 25 \\
 &     & & & &\\
 & W  & 83 & 30 & 38& 64\\
 &     & & & &\\
\hline
\multicolumn{5}{l}{Total points = 6253; $PC$=77.7\%}; $EBD$=1.9\% \\
\multicolumn{5}{l}{$POD(N)$=92.8\%; $POD(E)$=40.1\%} \\
\multicolumn{5}{l}{$POD(S)$=69.5\%; $POD(W)$=27.7\%} \\
\end{tabular}
\end{center}
\end{table}
In Table~\ref{tab:ct_cp_wd}, the samples are divided in the following 4 quadrants: (1) wind blowing from the direction between 0$^{\circ}$ and 90$^{\circ}$ (NE winds);
 (2) wind blowing from the direction between 90$^{\circ}$ and 180$^{\circ}$ (SE winds); (3) wind blowing from the direction between 180$^{\circ}$ and 270$^{\circ}$ 
(SW winds); and (4) wind blowing from the direction between 
270$^{\circ}$ and 360$^{\circ}$ (NW winds).
These quadrants are reported in black in Fig.~\ref{fig:wind_circle}.
$PC$ is equal to 67.7\%.
The probability of detection of the wind between 0$^{\circ}$ and 90$^{\circ}$ (NE winds) is around 82\%;
between 90$^{\circ}$ and 180$^{\circ}$ (SE winds), it is 65.5\%; between 180$^{\circ}$ and 270$^{\circ}$ (SW winds) it is 27.9\%;
between 270$^{\circ}$ and 360$^{\circ}$ (NW winds), it is around 42\%.
The model performance is very good in the quadrant in which the wind speed flows more frequently (between 0$^{\circ}$ and 90$^{\circ}$ with 82\%). 
The high frequence of occurence of the NE winds at Cerro Paranal is confirmed by the windrose
of Figure~\ref{fig:windrose} (left), between 1997 and 1999.
The worst $POD$ is found for the third quadrant (27.9\%), but this corresponds to only less than
4\% of the observed wind total sample at Cerro Paranal. 
This low frequence of occurence is also confirmed by the windrose of Figure~\ref{fig:windrose} (left).
In the other quadrants the model performance is, in most cases, well above the random case (25\%). 
We note that the low frequency in the quadrant [180$^{\circ}$, 270$^{\circ}$]
is confirmed by the climatology, and is not limited to our sample of 129 nights. 
If we divide now the sample in 4 different intervals (N, E, S and W, i.e. rotated of 45$^{\circ}$ with respect to the previous one 
in order to better constrain the angular sectors), we observe that the results remains very similar.
In Table~\ref{tab:ct_cp_wd2}, this time the samples 
are divided in the following 4 quadrants: (1) wind blowing from the direction between -45$^{\circ}$ and 45$^{\circ}$ (N winds);
 (2) wind blowing from the direction between 45$^{\circ}$ and 135$^{\circ}$ (E winds); (3) wind blowing from the direction between 135$^{\circ}$ and 225$^{\circ}$ 
(S winds); and (4) wind blowing from the direction between
225$^{\circ}$ and 315$^{\circ}$ (W winds).
These quadrants are reported in red in Fig.~\ref{fig:wind_circle}.
We notice that
$PC$ remain high, as expected, with a value of 77.7\%.
$POD$ are very good (as high as 92.8\% for $POD(N)$),
except for $POD(W)$ (less than 28\%).
In all cases, $EBD$ is very low, always inferior to 2.8\%, which confirms the goodness of the prediction by the model. \\
$POD(SW)$ (Tables~\ref{tab:ct_cp_wd}) and $POD(W)$ (Tables~\ref{tab:ct_cp_wd2}) are the only $POD$s giving 
values close than the random case, we can deduce that the wind blowing between 225$^{\circ}$ and 270$^{\circ}$ (hatched section 
of Fig.~\ref{fig:wind_circle}) is reconstructed with the most apparent difficulty. 
However, as previously mentioned, the observed frequency of these events is very low. \\
The results for the probability of detection of the wind direction by Meso-NH are summarized in Figure~\ref{fig:windrose} (right).
%
\section{Conclusions}
\label{sec9}
In this study, we have quantified the quality of the predictions of the absolute temperature, the wind speed and the wind direction in the surface layer 0-30~m at the ESO site of Cerro Paranal, 
by the atmospherical non-hydrostatical mesoscale model Meso-NH. 
First, we identified the bias and RMSE between model and observations. 
With respect to the study of \citet{lascaux2013}, the sample is more homogeneous (nights distributed on different periods of the year), and richer (129 nights instead of 20).
This permitted us to obtain very reliable and robust values for the bias, the RMSE and the bias-corrected RMSE $\sigma$.\\
To quantify the quality of the model predictions, simulations of 129 nights (distributed between 2007 and 2011) have been performed.
In addition to this, we also have constructed contingency tables and analyzed different parameters deduced from these tables, more precisely
the $PC$ (percent of correct detection), the $POD$ (probability of detection of a single event), and the $EBD$ (percent of extremely bad detections). 
These contingency tables (and the associated statistical parameters) 
allowed us to quantify the performance of the Meso-NH model above Cerro Paranal.\\ \\
For the absolute temperature, the percent of correct detection ($PC$) computed from 3$\times$3 contingency tables is excellent.
It is around 75\% range depending on the height above the ground (2~m and 30~m).
$POD$ are mostly larger than 66\%. It is 
and as good as 93.7\% at 30~m for the detection of temperature inferior to 11.5$^{\circ}$C.  The worst performances are observed in detecting the temperature in the narrow range [11.5$^{\circ}$,13.5$^{\circ}$] at 30~m (POD = 55.2\%). At every levels, the $EBD$ (extremely bad detections) 
are almost all equal to 0\%.
In all cases one can note that the $PC$ and the $POD$ are well greater than 33$\%$ (typical value for a random distribution for a 3$\times$3 contingency table). \\ \\
For the wind speed, the percent of correct detection ($PC$) computed from 3$\times$3 contingency tables is good (around 60\%, when the best configurations, $\Delta$X=100~m, 
is used).
The probability to detect the wind velocity in the three sub-samples limited by the tertiles of the cumulative distribution is good too.
The strongest wind are especially well detected by the model: $POD_3$ is equal to 74\% (at 10~m) and is equal to 79\% (at 30~m).\\
In all cases the $PC$ and the $POD$ are greater than 33$\%$, typical value for a random distribution. 
We also proved that the configuration with the horizontal resolution of 100~m in the innermost domain provides a much better model performances in reconstructing strong wind speed. \\ \\
For the wind direction, differently from the temperature and the wind speed, we have analyzed 4$\times$4 contingency tables.
If we consider the separation in the four categories (NE, SE, SW, NW), the $PC$ is 67.5\%. 
If we consider the separation in the four categories (N, E, S, W), the $PC$ is 77.7\%.
Both values are much better than for a random forecast ($PC$=25\%) of a 4$\times$4 contingency table.
The POD of the wind is very good in the quadrants from which the wind flows more frequently (N and NE) with PODs of the order of 92.8\% (N) and 81.7\% (NE). 
The POD is not satisfactory ($\sim$ 27\%) only in the narrow angular sector [225$^{\circ}$, 270$^{\circ}$]. 
However the occurrence of wind blowing from SW is the less probable (only 4\% of the time for the observed sample of 129 nights). 
The impact of the model performances on this case can be considered therefore negligible.\\ \\
Concerning the relative humidity, we have shown (in the appendix) that the model is able to reconstruct the high RH (as high as 80\% or more), condition for which the dome 
is closed. 
\\ \\
We therefore conclude that the model performances in reconstructing the absolute temperature, the wind speed and direction in the surface layer [0,30m] above the ground is very 
satisfactory 
and results of this study guarantee a concrete practical advantage from  the implementation of an automatic system for the forecasts of these parameters.
\section*{Acknowledgements}
Meteorological data-set from the AWS and mast at Cerro Paranal are from ESO Astronomical Site Monitor (ASM - Doc.N. VLT-MAN-ESO-17440-1773). 
We are very grateful  
to the ESO Board of MOSE (Marc Sarazin, Pierre-Yves Madec, Florian Kerber and Harald Kuntschner) for their constant support to this study. 
We acknowledge M. Sarazin and F. Kerber for providing us the ESO data-set used in this study.  
A great part of simulations are run on the HPCF cluster of the European Centre for Medium Weather Forecasts (ECMWF) - Project SPITFOT. 
This study is co-funded by the ESO contract: E-SOW-ESO-245-0933.

\clearpage
\appendix
\section{Cumulative distributions}
We report in this section the cumulative distributions of bias, RMSE and $\sigma$ for all parameters and at all levels at Cerro Paranal, 
for the sample pf 129 nights. 
\begin{figure*}
\begin{center}
\begin{tabular}{c}
\includegraphics[width=0.95\textwidth]{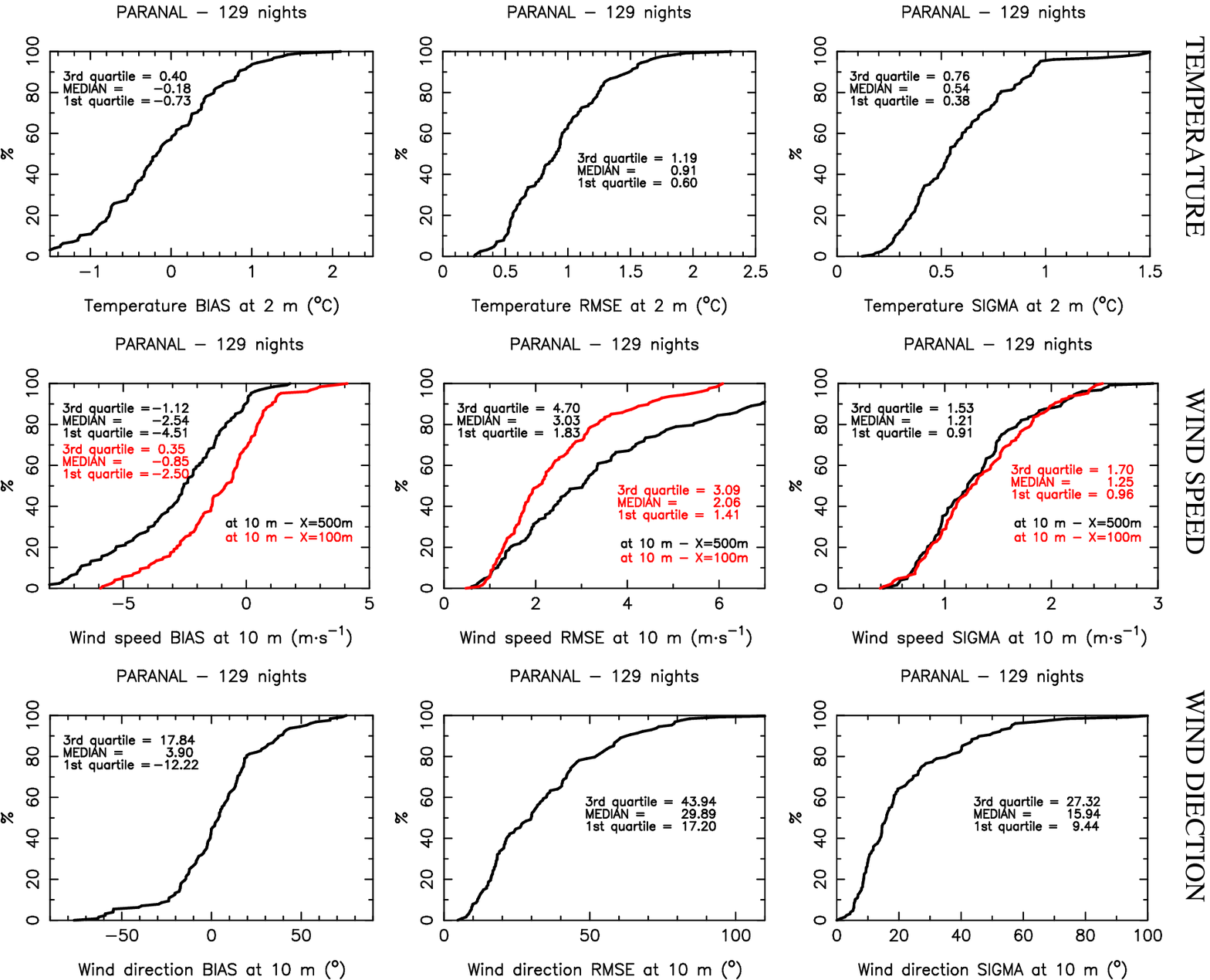}\\
\end{tabular}
\end{center}
\caption{\label{fig:cumdist} Cumulative distribution of the single nights bias, RMSE and bias-corrected RMSE for the absolute temperature (top row),
the wind speed (middle row, $\Delta$X~=~500~m and $\Delta$X~=~100~m configurations) and the wind direction (bottom row), at 2~m (temperature) and
at 10~m (wind speed and direction).}
\end{figure*}
\begin{figure*}
\begin{center}
\begin{tabular}{c}
\includegraphics[width=0.95\textwidth]{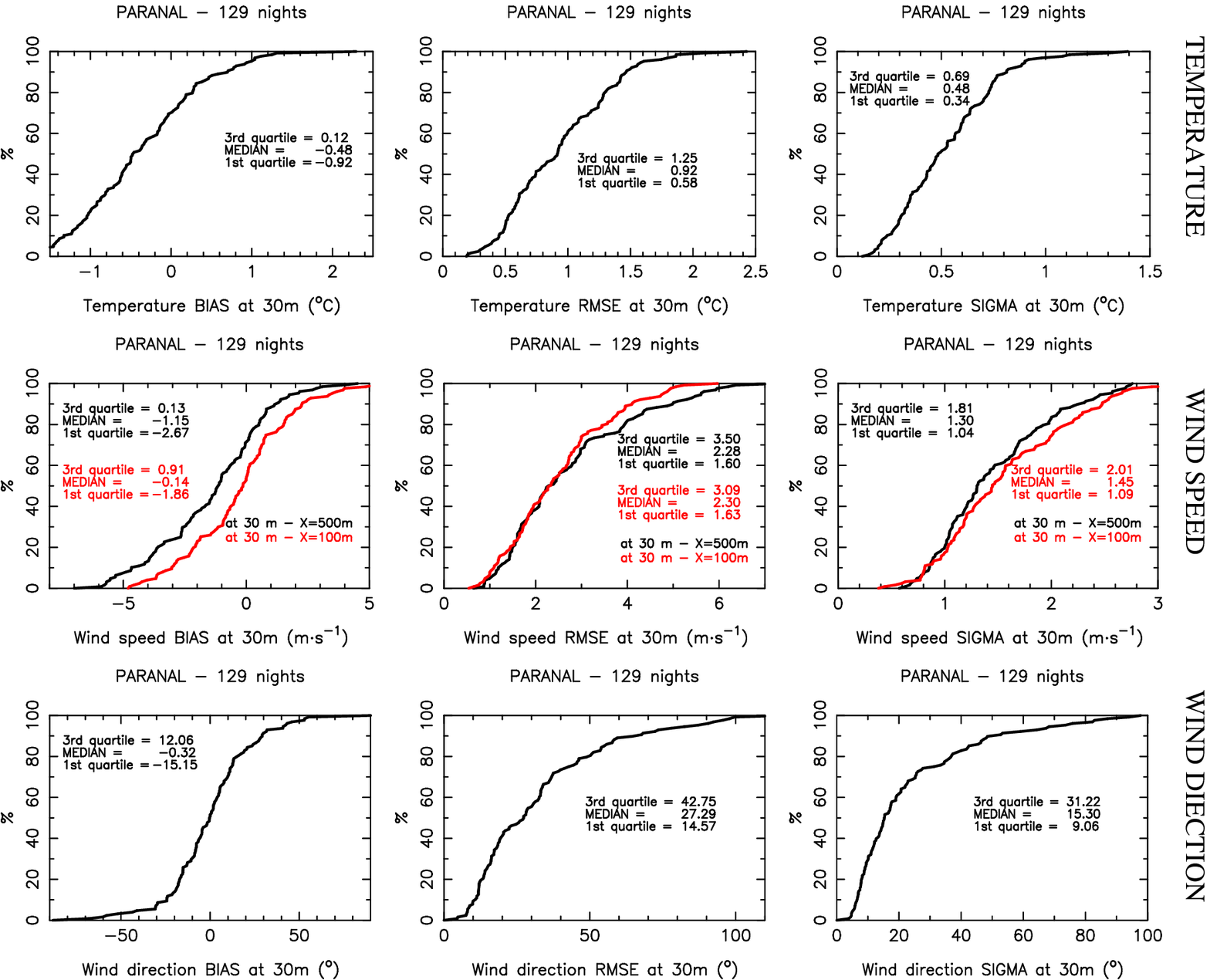}\\
\end{tabular}
\end{center}
\caption{\label{fig:cumdist2} Cumulative distribution of the single nights bias, RMSE and bias-corrected RMSE for the absolute temperature (top row),
the wind speed (middle row, $\Delta$X~=~500~m and $\Delta$X~=~100~m configurations) and the wind direction (bottom row), at 30~m.}
\end{figure*}
\clearpage
\begin{figure*}
\begin{center}
\begin{tabular}{c}
\includegraphics{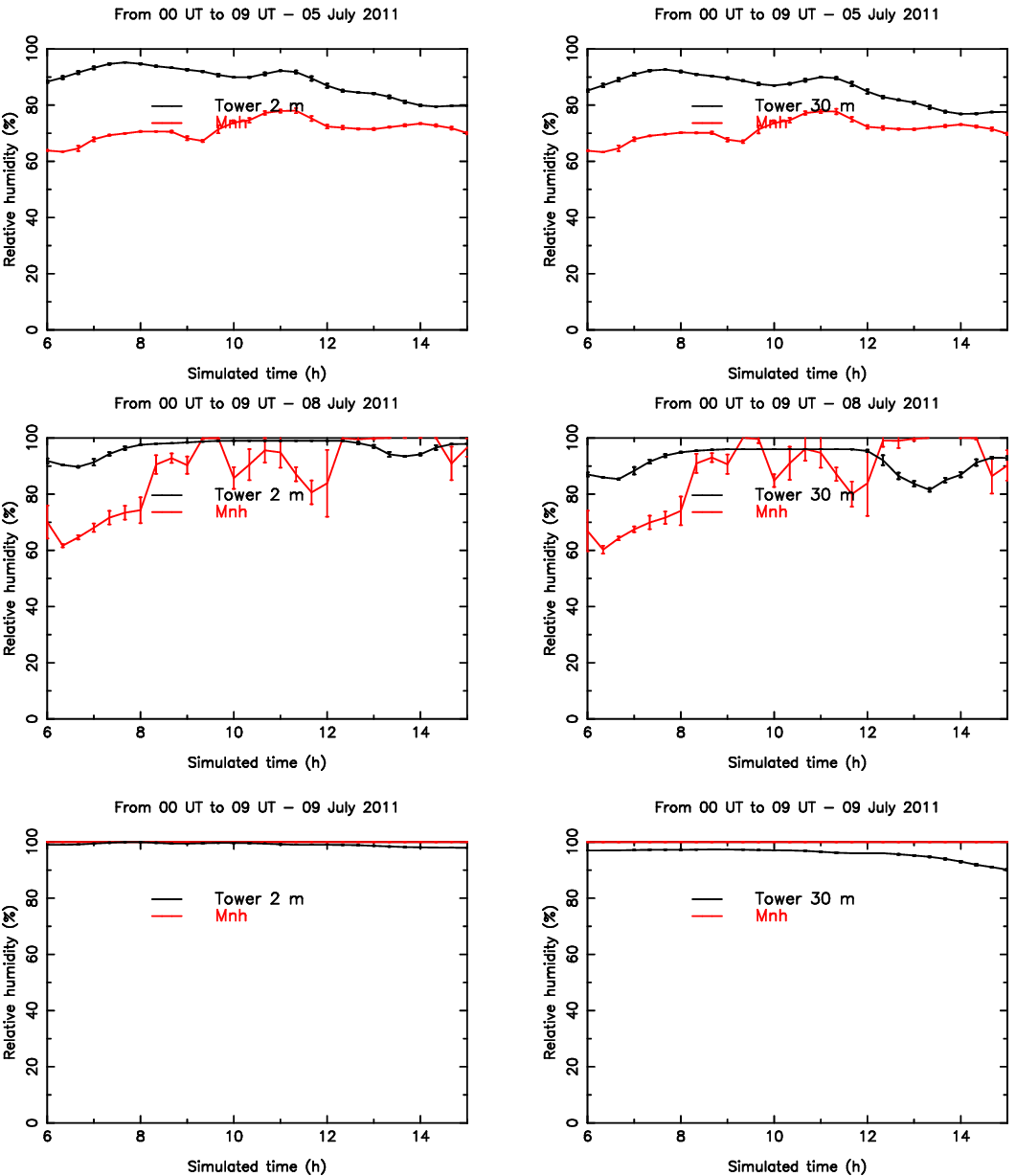}
\end{tabular}
\end{center}
\caption[appendix_rh]{\label{fig:appendix_rh} Temporal evolutions during the night of the relative humidity (in \%) at 2~m (left)
and 30~m (right) at Cerro Paranal, for 3 nights in July 2011. In black: the observations; in red: the model outputs.}
\end{figure*}
\begin{figure*}
\begin{center}
\begin{tabular}{c}
\includegraphics{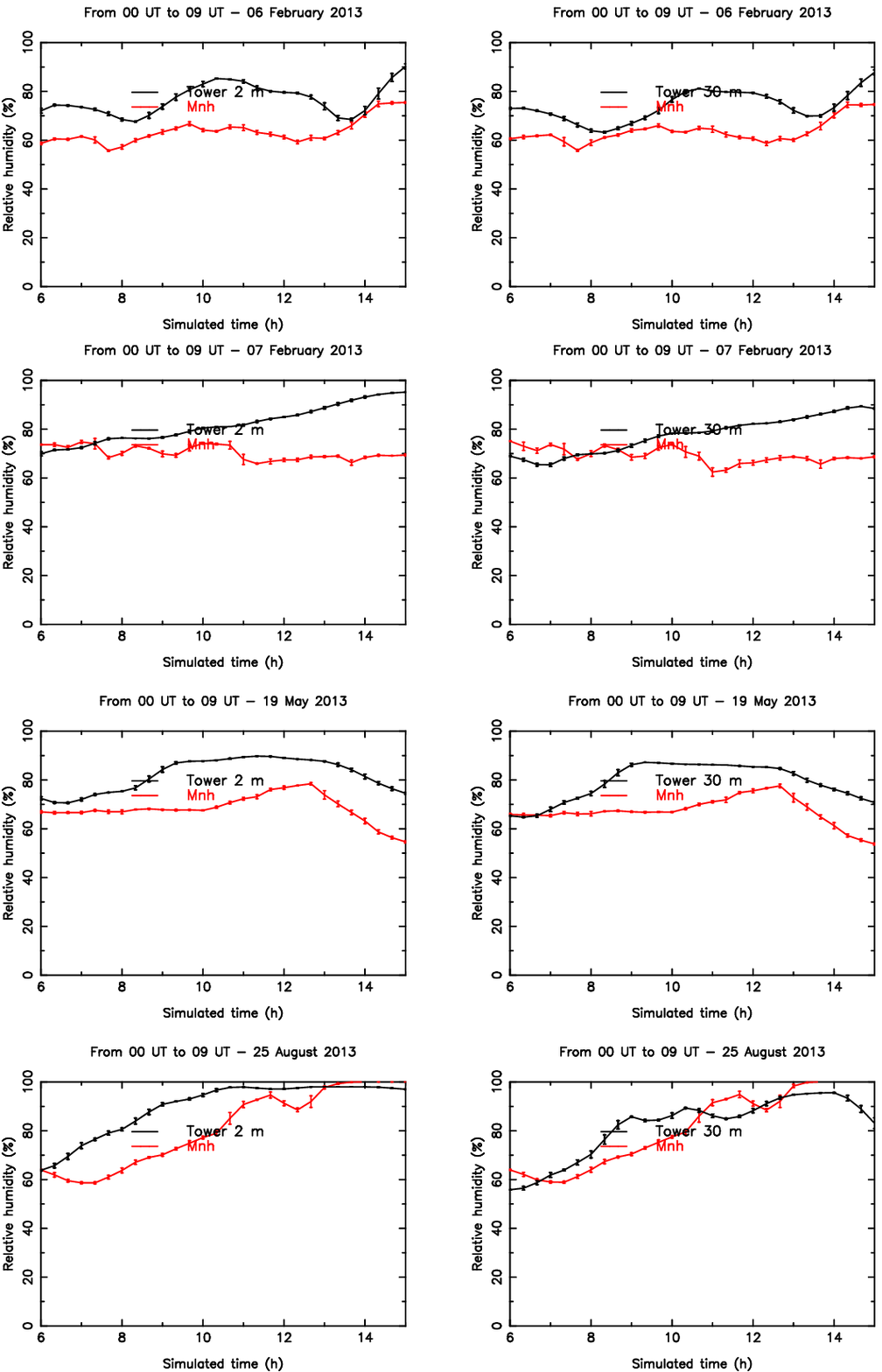}
\end{tabular}
\end{center}
\caption[appendix_rh_2013]{\label{fig:appendix_rh_2013} Temporal evolutions during the night of the relative humidity (in \%) at 2~m (left)
and 30~m (right) at Cerro Paranal, for 4 nights in 2013. In black: the observations; in red: the model outputs.}
\end{figure*}

\section{Relative humidity}
\noindent
We report in this section the forecasted temporal evolutions for 3 nights in 2011, and 4 nights in 2013, of the relative humidity.
These nights are all the nights of 2011 and 2013, in which a relative humidity (RH) larger than 80\% has been observed. 
This is the unique interesting condition to be studied for astronomical application in this region for these wavelengths. As already remembered, 
in this region the RH is mainly of the order of 20-30 \% and measurements for such a low values are not necessarily reliable.
Figure~\ref{fig:appendix_rh} displays the temporal evolution of the relative humidity for these nights.
The forecasted relative humidity is, in these cases, very close to the observed one. In two nights over three in 2011 the forecasted RH is moreover
larger than 80\% as well as the observed one. In one night (5 July 2011), the forecasted RH is slightly inferior ($\sim$ 70\%) but still very high. 
The four night of 2013 with a high observed RH are well reproduced too by the model (predicted values always between 60\% and 100\%) even if a small dry bias 
is present.
This means that apparently, the model succeeds in predicting these critical high values, even though the frequency of occurrence of such an event is very low. 
Practically, the astronomer needs to know if the humidity during the night will 
be higher than 80\%, a threshold above which the dome must be closed, and thus observations are not allowed.
Our results indicate that the model can give a positive forecast of the night relative humidity useful for astronomers. Of course it would be of interest to 
identify a richer sample of nights with relative humidity as high as 80\% across several years and study the behavior of the model on a richer sample.
This might permit us to have a more robust conclusions on this specific subject.

\label{lastpage}
\end{document}